%% file: cop.tex
\documentclass[12pt]{article}
\usepackage{epsfig}

\input econfmacros.tex
\def\beq{\begin{equation}}
\def\enq{\end{equation}}
\def\ba{\begin{array}}
\def\ea{\end{array}}
\def\Mesz{M\'esz\'aros}
\def\siml{\lower4pt \hbox{$\buildrel < \over \sim$}}
\def\simg{\lower4pt \hbox{$\buildrel > \over \sim$}}
\def\citen{\cite}

\newcommand{\boxsize}{0.89\textwidth}

\def\Title#1{\begin{center} {\Large {\bf #1} } \end{center}}

\begin{document}

\Title{High Energy Photons, Neutrinos and Gravitational Waves from 
Gamma-Ray Bursts}

\bigskip\bigskip

%+\addtocontents{toc}{{\it P. M\'esz\'aros et al.}}
%+\label{ReggianoStart}

\begin{raggedright}

{\it P. M\'esz\'aros, S. Kobayashi, S. Razzaque \& B. Zhang\index{M\'esz\'aros, P., et al.}\\
Dept. of Astronomy \& Astrophysics and Dept. of Physics\\
Pennsylvania State University\\
University Park, PA 16802, USA}
\bigskip\bigskip
\end{raggedright}

\section{Introduction}

A new era in Gamma-ray Burst (GRB) research opened in 1991 with the launch of 
the Compton Gamma-Ray Observatory (CGRO), whose ground-breaking results have been 
summarized in \cite{fm95}.  The most significant results came from the all-sky 
survey by
the Burst and Transient Experiment (BATSE) on CGRO, which recorded over 2700
bursts, complemented by data from the OSSE, Comptel and EGRET experiments,
in the 20 keV - 1 MeV, 0.1-10 MeV, 1-30 MeV and 20 MeV - 30 GeV gamma-ray
bands, respectively. The next major advance was ushered in by  the Italian-Dutch 
satellite Beppo-SAX in 1997, which obtained the first high resolution 0.2-10 keV 
x-ray images \cite{cos97} of the fading afterglow of a burst, GRB 970228, a
phenomenon which had been previously theoretically predicted. This was
followed by an increasing list of burst afterglow detections with Beppo-SAX.
These X-ray detections, after a 4-6 hour delay needed for data processing,
led to arc-minute accuracy positions, which finally made possible the optical
detection and the follow-up of the GRB afterglows at longer wavelengths (e.g.
\cite{jvp97,fra97}). This paved the way for the measurement of redshift
distances, the identification of candidate host galaxies, and the confirmation
that they were at cosmological distances \cite{metz97,kul99b}, etc.
In the last year, the HETE-2 spacecraft \cite{hete-2} has been discovering
bursts at the rate of $\sim$ 2/month, in the 0.5-400 keV range. Besides
classical bursts, a number of these have been in the sub-class of X-ray
flashes, first identified with Beppo-SAX \cite{heise00}, or X-ray rich
bursts \cite{kippen02}.

Another remarkable observation \cite{ak99} was that of a prompt and
extremely bright ($m_v\sim 9$) optical flash in the burst GRB 990123, 15
seconds after the GRB started (and while it was still going on). A prompt
multi-wavelength flash, contemporaneous with the $\gamma$-ray emission and
reaching such optical magnitude levels is an expected consequence of the
reverse component of external shocks \cite{mr93b,mr97a}. The prompt optical
flash of 990123 is generally interpreted \cite{sp99,mr99} as the radiation
from a reverse external shock (although prompt optical flashes can be
expected from both an external reverse shock and from an internal shock
\cite{mr97a}). The behavior of the reverse shock is more sensitive than the 
forward shock on the details of the shock physics
\cite{kobayashi00}. This has become apparent with the recent prompt flashes
seen from ground-based optical follow-up of two other bursts, GRB 021004 and
GRB 021211, \cite{fox03a,fox03b,li03}, made possible by HETE-2 spacecraft 
position alerts.
In the near future, the prospects look bright for detecting 100-150 bursts
per year with a new multi-wavelength GRB follow-up spacecraft  called
{\it Swift} \cite{swift}, which is due for launch in December 2003 and is
equipped with $\gamma$-ray, x-ray and optical detectors. Predictions of the
dependence of spectra and lightcurves on the initial bulk Lorentz factor,
magnetic energy content, etc. \cite{zhangkobmes03} opens up the interesting
possibility of charting in much greater detail the shock physics and the
dynamics of the relativistic outflow in GRB.

The large bulk of the information on GRB has come from electromagnetic signals
at MeV and lower energies, except for a handful of bursts detected at higher
photon energies with SMM \cite{smm}, with the EGRET, OSSE and Comptel
experiments on CGRO \cite{fm95}, and more recently INTEGRAL \cite{integral}.
However, new missions are planned for ultra-high energies (UHE). One of these,
aimed at the 20 MeV - 300 GeV as well as as 10 keV - 30 MeV range, is GLAST 
\cite{glast00}, due for launch in 2006. In addition, several new ground-based 
air and water Cherenkov telescopes as well as solar array facilities are 
coming or will shortly come on-line, sensitive in the 0.2-10 TeV range, with 
GRB as one possible class of targets \cite{weekes00}. 
Furthermore, there are several other, as yet unconfirmed,
but potentially interesting observing channels for GRBs, relating to the
baryonic  component of the outflow, the shock physics and the progenitor
collapse dynamics. There is the prospect of measuring ultra-high energy
cosmic rays from GRB with the Auger array \cite{auger}; TeV and higher energy
neutrinos with large, cubic kilometer ice or water Cherenkov telescopes such
as ICECUBE or ANTARES \cite{halzen00}; and measuring gravitational
wave signals accompanying the GRB formation with LIGO and similar arrays 
\cite{finn00}. We discuss below some of the theoretical expectations for 
the GRB ultra-high energy (UHE) $\simg$ GeV electromagnetic signals, the 
UHE neutrino  signals, and the gravitational wave signals.

\section{UHE photons from GRB}

Ultra-high energy emission, in the range of GeV and harder, is expected from
from electron inverse Compton in external shocks \cite{mpr94} as well as
from internal shocks \cite{pm96} in the prompt phase. The combination of
prompt MeV radiation from internal shocks and a more prolonged GeV IC
component for external shocks \cite{mr94} is a likely explanation for
the delayed GeV emission seen in some GRB  \cite{hur94}.
(For an alternative invoking photomeson processes from ejecta protons
impacting a nearby binary stellar companion see \cite{katz94}).
The GeV photon emission from the long-term IC component in external afterglow
shocks has been considered by \cite{zhang01,sariesin01,derishev01}. The IC GeV
photon component is likely to be significantly more important \cite{zhang01}
than a possible proton synchrotron or electron synchrotron component at these
energies. Another possible contributor at these energies may be $\pi^0$ decay
from $p\gamma$ interactions between shock-accelerated protons and MeV or
other photons in the GRB shock region \cite{boetder98,totani99,fragile02}.
However, under the conservative assumption that the relativistic proton
energy does not exceed the energy in relativistic electrons or in $\gamma$-rays,
and that the proton spectral index is -2.2 instead of -2, both the proton
synchrotron and the $p\gamma$ components can be shown to be substantially
less important at GeV-TeV than the IC component \cite{zhang01}.
Another GeV photon component is expected from the fact that in a
baryonic GRB outflow neutrons are likely to be present, and when these
decouple from the protons, before any shocks occur, $pn$ inelastic
collisions will lead to pions, including $\pi^0$, resulting in UHE
photons which cascade down to the GeV range \cite{derishev99,bahmesz00}.
The final GeV spectrum results from a complex cascade, but a rough estimate
indicates that 1-10 GeV flux should be detectable with GLAST for bursts
at $z\siml 0.1$ \cite{bahmesz00}.

%%%%%%%%%%%%%%%%%%%%%%%%%%%%%%%%%%%%%%%%%%%%%%%%%%%%%%%%%%%%%%%%%%%%%%%%%
\begin{figure}[htb]
\begin{center}
\begin{minipage}[t]{0.45 \textwidth}
\epsfxsize=\boxsize
\epsfbox{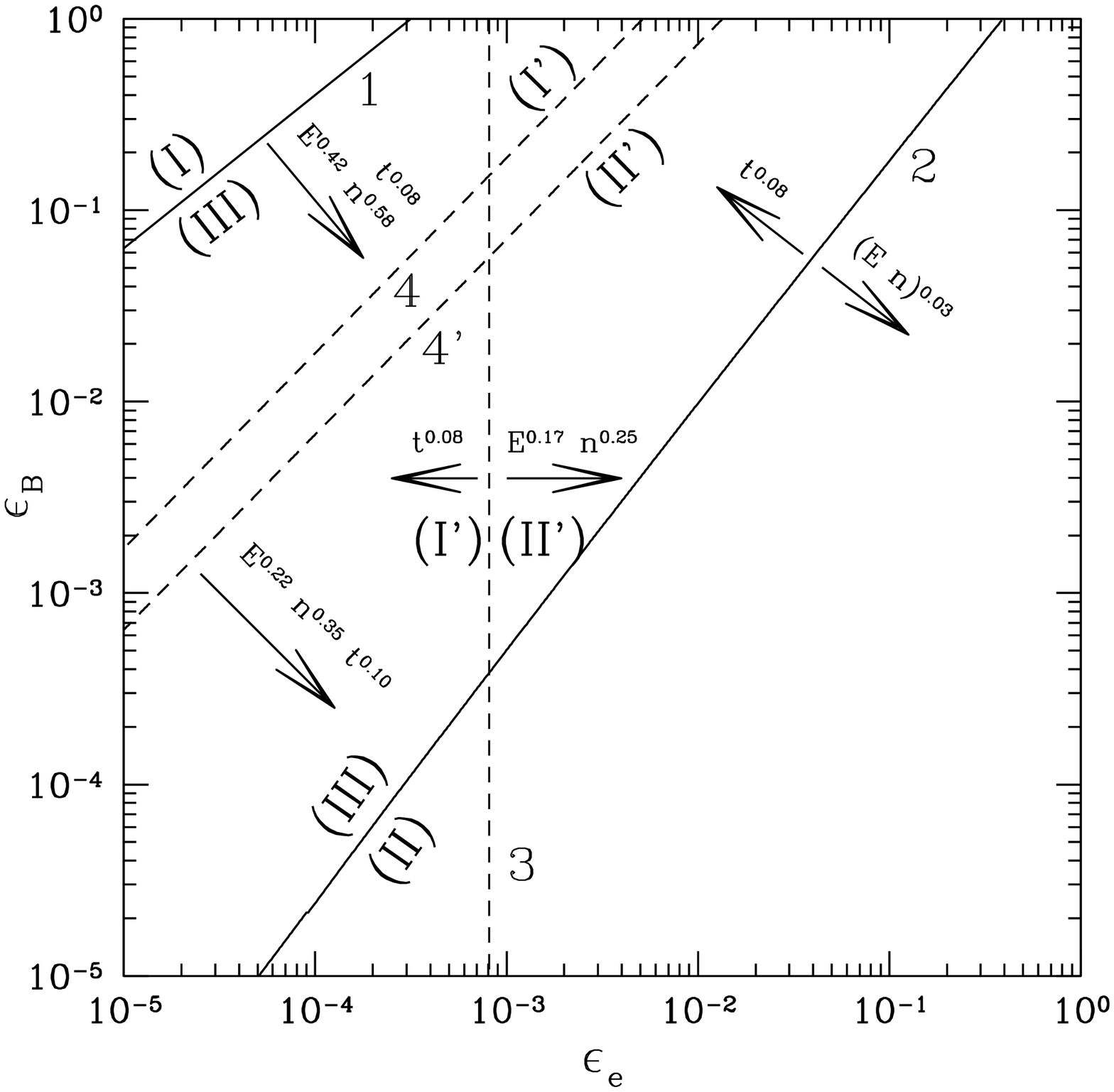}
\end{minipage}
\hspace{3mm}
\begin{minipage}[t]{0.45\textwidth}
\epsfxsize=\boxsize
\epsfbox{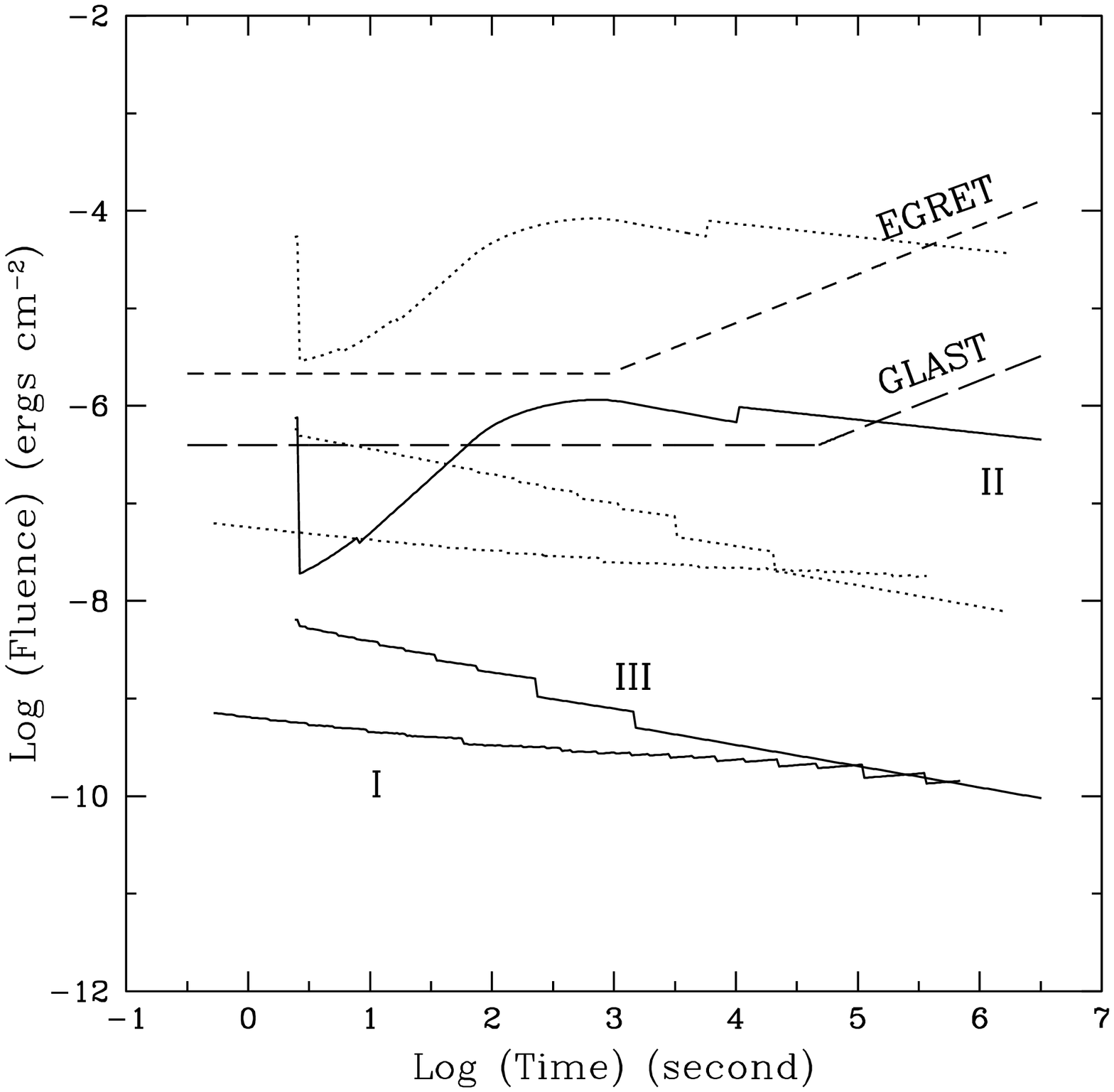}
\end{minipage}
\caption{{\it Left panel}: Parameter regimes for GRB GeV emission dominated by proton
synchrotron (I), electron synchrotron (III) and electron inverse Compton (II).
{\it Right panel}: GeV lightcurves for the three types of bursts in the range
400 MeV - 200 GeV. The solid curves indicate bursts in regimes I (p-sy), II (e-IC)
and III (e-sy), at a distance $z=1$. The three dotted unmarked curves are the same
but located at $z=0.1$. Also shown are the sensitivity curves for {\em EGRET} and 
{\em GLAST}}.  
\label{fig:uhegamregimes}
\end{center}
\end{figure}
%%%%%%%%%%%%%%%%%%%%%%%%%%%%%%%%%%%%%%%%%%%%%%%%%%%%%%%%%%%%%%%%%%%%%%%%%%%

In these models, due to the high photon densities implied by GRB models,
$\gamma\gamma$ absorption within the GRB emission region must be taken into
account (see also \cite{baring00,lithwick01}). So far, these calculations
have been largely analytical. One interesting result is that the observation
of photons up to a certain energy, say 10-20 GeV with EGRET, puts a lower
limit on the bulk Lorentz factor of the outflow, from the fact that the
compactness parameter (optical depth to $\gamma\gamma$) is directly
proportional to the comoving photon density, and both this as well as the
energy of the photons depend on the bulk Lorentz factor. This has been used
by \cite{lithwick01} to estimate lower limits on $\Gamma$ in the range 300-600
for a number of specific bursts observed with EGRET.

At higher energies, a tentative $\simg 0.1$ TeV detection at the $3\sigma$
level of GRB970417a has been reported with the water Cherenkov detector
Milagrito \cite{atk00}.  Another possible TeV detection \cite{poirier03}
of GRB971110 has been reported with the GRAND array, at the $2.7\sigma$ level.
Stacking of data from the TIBET array for a large number of GRB time windows
has led to an estimate of a $\sim 7\sigma$ composite detection significance
\cite{amenomori01}.
Better sensitivity is expected from the upgraded larger version of
MILAGRO, as well as from atmospheric Cherenkov telescopes under
construction such as VERITAS, HESS, MAGIC and CANGAROO-III \cite{weekes00}.
However, GRB detections in the TeV range are expected only for rare nearby
events, since at this energy the mean free path against $\gamma\gamma$
absorption on the diffuse IR photon background is $\sim$ few hundred Mpc
\cite{coppi97,ste00b}.  The mean free path is much larger at GeV energies,
and based on the handful of GRB reported in this range with EGRET,
several hundred should be detectable with large area
space-based detectors such as GLAST \cite{glast00,zhang01}.

\section{UHE neutrinos from GRB}

While stellar scenarios for GRB will result in large thermal ($\sim 10-30$
MeV) neutrino luminosities comparable to those in supernovae, at typical
redshifts $z\sim 1$ these are extremely hard to detect due to the low cross
sections at these energies. However, the neutrino detection  cross section
increases with energy, and near TeV energies there are realistic chances
for detection \cite{halzen00} with cubic kilometer ice or water Cherenkov
detectors, such as the planned ICECUBE or ANTARES.

A mechanism leading to higher (GeV) energy neutrinos in GRB is inelastic
nuclear collisions. Proton-neutron inelastic collisions are expected, at
much lower radii than where shocks occur, due to the decoupling of neutrons
and protons in the fireball or jet \cite{der99a}. If the fireball has a
substantial neutron/proton ratio, as expected in most GRB progenitors, the
collisions become inelastic and their rate peaks at when the nuclear scattering
time becomes comparable to the expansion time.  This occurs when the $n$ and
$p$ fluids decouple, their relative drift velocity becoming comparable to $c$,
which is easier due to the lack of charge of the neutrons. Inelastic $n,p$
collisions  then lead to charged pions and GeV muon and electron neutrinos
\cite{bm00}. The early decoupling and saturation of the $n$ also leads to
a somewhat higher final $p$ Lorentz factor \cite{der99a,bm00,fuller00,pruet03},
implying a possible relation between the $n/p$ ratio and the observable fireball
dynamics, relevant for the progenitor question. Fusion and photodisociation
processes also play a role in such effects \cite{lemoine02,belobor02}.
Inelastic $p,n$ collisions leading to neutrinos can also occur
in fireball outflows with transverse inhomogeneities in the bulk Lorentz factor,
where the $n$ can drift sideways into regions of different bulk velocity flow,
or in situations where internal shocks involving $n$ and $p$ occur close to the
saturation radius or below the photon photosphere \cite{mr00a}.  The typical
$n,p$ neutrino energies are in the 1-10 GeV range, which could be detectable in
coincidence with observed GRBs for a sufficiently close  photo-tube spacing
in km$^3$ detectors such as ICECUBE \cite{bm00}.

Neutrinos at energies $\simg$ 100 TeV  are expected from $p,\gamma$
photo-pion interactions between relativistic protons accelerated in the
fireball internal or external shocks. A high collision rate is ensured here
by the large density of photons in the fireball shocks. The most straightforward
is the interaction between MeV photons produced by radiation from electrons
accelerated in internal shocks and relativistic protons accelerated by the
same shocks \cite{wb97}, leading to charged pions, muons and neutrinos. This
$p,\gamma$ reaction peaks at the energy threshold for the photo-meson $\Delta$
resonance in the fluid frame moving with $\gamma$, or
$ \eps_p\eps_\gamma \simg 0.2 \hbox{GeV}^2 \gamma^2$.
For observed 1 MeV photons this implies
$\simg 10^{16}$ eV protons, and neutrinos with $\sim 5\%$ of that energy,
$\eps_\nu\simg 10^{14}$ eV in the observer frame. Above this threshold, the
fraction of the proton energy lost to pions is $\sim 20\%$ for typical fireball
parameters, and the typical spectrum of neutrino energy per decade is flat,
$\eps_\nu^2 \Phi_\nu \sim$ constant. Synchrotron and adiabatic losses limit the
muon lifetimes \cite{rachenm98}, leading to a suppression of the neutrino flux
above $\eps_\nu \sim 10^{16}$ eV. In external shocks another copious source of
targets are the O/UV photons in the afterglow reverse shock (e.g. as deduced from
the GRB 990123 prompt flash of \citen{ak99}). In this case the resonance condition
implies higher energy protons, leading to neutrinos of $10^{17}-10^{19}$ eV
\cite{wb99a,vie98a}. These neutrino fluxes are expected to be detectable above
the atmospheric neutrino background with the planned cubic kilometer ICECUBE
detector \cite{halzen00}. Useful limits to their total contribution to
the diffuse ultra-high energy neutrino flux can be derived from observed cosmic
ray and diffuse gamma-ray fluxes \cite{wb99b,bw01,mann01}.
While the $p,\gamma$ interactions leading to $\simg 100$
TeV energy neutrinos provide a direct probe of the internal shock acceleration
process, as well as of the MeV photon density associated with them, the $\simg
10$ PeV neutrinos would probe the reverse external shocks, as well as the photon
energies and energy density there.

The most intense neutrino signals may be due to $p,\gamma$ interactions
occurring {\it inside} collapsars while the jet is still burrowing its way out of
the star \cite{mw01}, before it has had a chance to break through (or not) the
stellar envelope to produce a GRB outside of it.  While still inside the star, the
buried jet produces thermal X-rays at $\sim 1$ keV which interact with $\simg 10^5$
GeV protons which could be accelerated in internal shocks occurring well inside
the jet/stellar envelope terminal shock, producing $\simg$ few TeV energy
neutrinos for tens of seconds, which penetrate the envelope.
The most conspicuous contribution of TeV $\nu_\mu$ is due to $pp,pn$ collisions
involving the buried jet relativistic protons and thermal nucleons in the jet
and the stellar envelope \cite{rmw03b}, while  $p\gamma$ is important at higher
energies $\simg 100 TeV$. At TeV energies the number of neutrinos is larger for
the same total energy output, which improves the detection statistics. The rare
bright, nearby or high $\gamma$ collapsars could occur at the rate of $\sim
10$/year, including both $\gamma$-ray bright GRBs (where the jet broke through the
envelope) and $\gamma$-ray dark events where the jet is choked (failed to break
through), and both such $\gamma$-bright and dark events could have a TeV neutrino
fluence of $\sim 10$/neutrinos/burst, detectable by ICECUBE in individual bursts.

%%%%%%%%%%%%%%%%%%%%%%%%%%%%%%%%%%%%%%%%%%%%%%%%%%%%%%%%%%%%%%%%%%%%%%%%%
\begin{figure}[htb]
\begin{center}
\begin{minipage}[t]{0.45 \textwidth}
\epsfxsize=\boxsize
\epsfysize=2.2in
\epsfbox{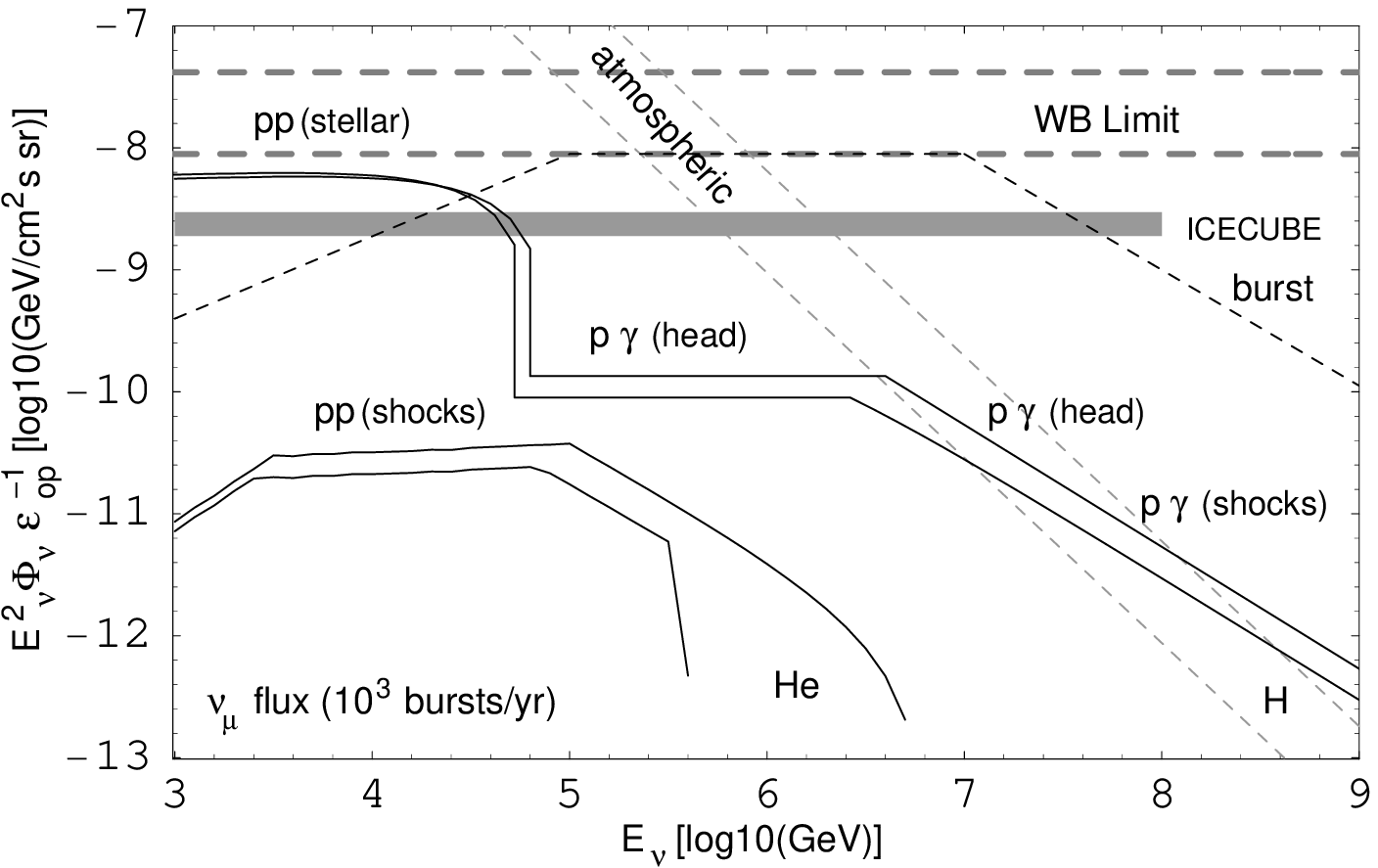}
\end{minipage}
\hspace{3mm}
\begin{minipage}[t]{0.45\textwidth}
\epsfxsize=\boxsize
\epsfysize=2.2in
\epsfbox{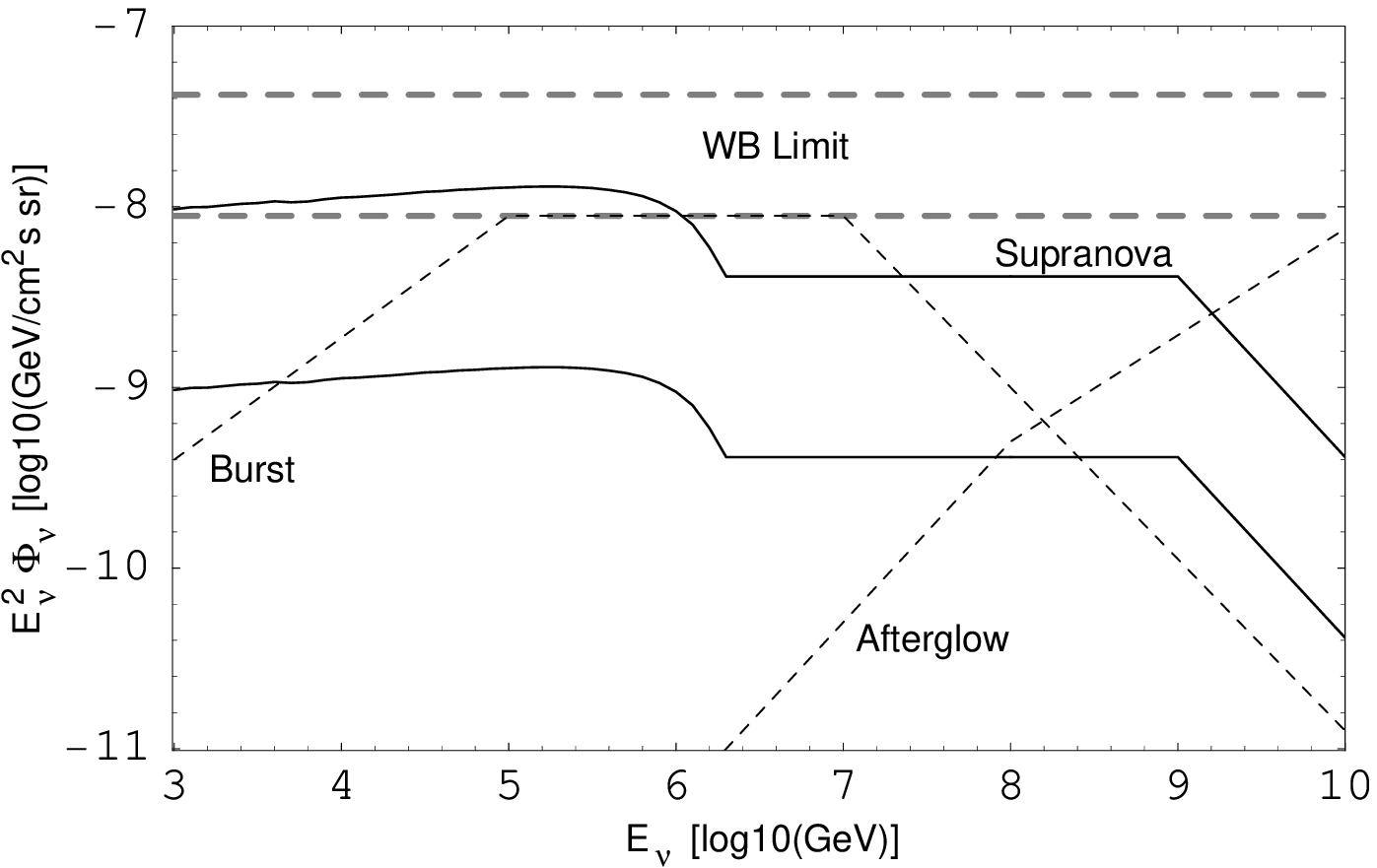}
\end{minipage}
\caption{{\it Left panel:} The diffuse muon-neutrino flux $E_{\nu}^2 \Phi_\nu
\varepsilon_{\rm op}^{-1}$ is shown as solid lines for $\nu_{\mu}$ from 
sub-stellar jet shocks in two GRB progenitor models, $r_{12.5}$ (H) and 
$r_{11}$ (He). These neutrinos arrive as precursors (10-100 s before) of 
$\gamma$-ray bright (electromagnetically detectable) bursts.  Also shown is the
diffuse neutrino flux arriving simultaneously with the $\gamma$-rays
from shocks outside the stellar surface in observed GRB (dark
short-dashed curve); the Waxman-Bahcall (WB) diffuse cosmic ray bound
(light long-dashed curves); and the atmospheric neutrino flux (light
short-dashed curves). For a hypothetical 100:1 ratio of $\gamma$-ray
dark (in which the jets do not emerge) to $\gamma$-ray bright collapses, 
the solid neutrino spectral curves would be 100 times higher.
{\it Right panel:} Diffuse neutrino flux ($E_{\nu}^2 \Phi_{\nu}$) from
post-supernova (supra-nova) models of GRBs (solid curves), assuming
that (top curve) all GRBs have an SNR shell, or (bottom) 10\% of all
GRBs have an SNR shell, and $10^{-1}$ of the fireball protons reach
the shells.  Long dashed lines correspond to the Waxman-Bahcall
cosmic-ray limit, short dashed curves are the diffuse $\nu$ flux from
GRB internal shocks and afterglows.}
\label{fig:nu_buriedjet}
\end{center}
\end{figure}
%%%%%%%%%%%%%%%%%%%%%%%%%%%%%%%%%%%%%%%%%%%%%%%%%%%%%%%%%%%%%%%%%%%%%%%%%%%

Recent reports of detection of X-ray lines from several GRB afterglows have
been interpreted \cite{piro00}, although not unambiguously, as providing
support for a version of the collapsar model in which a SN explosion
occurs weeks before the GRB (the ``supranova" model, \cite{vie98b}).
In the supranova scenario the supernova remnant (SNR) shell provides
nucleon targets for $pp$ interactions with protons accelerated in the
MHD wind of a pre-GRB pulsar \cite{gg02}, leading to a 10 TeV neutrino
precursor to the GRB (other nucleon targets from a stellar companion
disruption leading to $\pi^0$ decay GeV $\gamma$-rays were discussed
by \cite{katz94}). The SNR also provides additional target photons for
$p\gamma$ interactions \cite{gg02} with internal shock-accelerated
protons, resulting in $\sim 10^{16}$ eV neutrinos.
We have calculated the expected neutrino fluxes and muon event rates
from individual bursts as well as the diffuse contribution
from all bursts \cite{rmw03a}. The neutrinos
produced by $pp$ and $p\gamma$ interactions between GRB relativistic
protons and SNR shell target protons and photons will be contemporary
and of similar duration as the GRB electromagnetic event. The high
$pp$ optical depth of the shell also implies a moderately high average
Thomson optical depth $\tau_{\rm T} \propto t_d^{-2}$ of the shell,
dropping below unity after $\sim 100$ days. Large scale anisotropy as
well as clumpiness of the shell will result in a mixture of higher and
lower optical depth regions being observable simultaneously, as
required in the supranova interpretation of X-ray lines and photon
continua in some GRB afterglows \cite{piro00}.  Depending on the
fraction of GRB with SNR shells, the contributions of these to the GRB
diffuse neutrino flux has a $pp$ component which is relatively
stronger at TeV-PeV energies than the internal shock $p\gamma$
component of \cite{wb97}, and a shell $p\gamma$ component which is a
factor 1 (0.1) of the internal shock $p\gamma$ component
(Fig. \ref{fig:nu_buriedjet}) for a fraction 1 (0.1) of GRB with SNR
shells. Due to a higher synchrotron cooling break in the shell, at
$E_\nu\simg 10^{17}$ eV the shell component could compete with the
internal \cite{wb97} and afterglow \cite{wb99a} components.
Our $pp$ component is caused by internal shock-accelerated power-law
protons contemporaneous with the GRB event, differing from \cite{gg02}
who considered quasi-monoenergetic $\gamma_p\sim 10^{4.5}$ protons
from an MHD wind over $4\pi$ leading to a $\sim 10$ TeV neutrino
months-long precursor of the GRB. Our $p\gamma$ component arises from
the same GRB-contemporaneous internal shock protons interacting with
thermal 0.1 keV photons within the shell wall, whereas \cite{gg02}
consider such protons interacting with photons from the MHD wind
inside the shell cavity. These neutrino fluxes provide a test for
the supranova hypothesis, the predicted event rates being detectable
with kilometer scale planned Cherenkov detectors.

\section{Gravitational waves from GRB}

The leading models for the ultimate energy source of GRB are stellar collapse
or compact stellar mergers, and these are expected to be sources of 
gravitational waves (GWs). A time-integrated GW luminosity of the order of a solar 
rest mass ($\sim 10^{54}$ erg) is predicted from merging NS-NS and NS-BH models
\cite{kp93,ruja98,nakamura00}, while the luminosity from collapsar models is
more model-dependent, but expected to be lower (\citen{fryerwh01,muellergw01};
c.f. \citen{vanputten01}).
We have estimated the strains of gravitational waves from some of the
most widely discussed current GRB progenitor stellar systems \cite{km03}.
The expected detection rates of gravitational wave events with the Laser
Interferometric Gravitational Wave Observatory (LIGO) from compact binary
mergers, in coincidence with GRBs, has been estimated by \cite{finn00,finn03}.
If some fraction of GRBs are produced by double neutron star  or neutron star
-- black hole mergers, the gravitational wave chirp signal of the in-spiral
phase should be detectable by the advanced LIGO within one year, associated
with the GRB electromagnetic signal. We have also estimated the signals from
the black hole ring-down phase, as well as the possible contribution of a bar
configuration from gravitational instability in the accretion disk following
tidal disruption or infall in GRB scenarios. Other binary progenitor scenarios,
such black hole -- Helium star and black hole -- white dwarf merger
GRB progenitors are unlikely to be detectable, due to the low estimates
obtained for the maximum non-axisymmetrical perturbations.

%%%%%%%%%%%%%%%%%%%%%%%%%%%%%%%%%%%%%%%%%%%%%%%%%%%%%%%%%%%%%%%%%%%%%%%%%
\begin{figure}[htb]
\begin{center}
\begin{minipage}[t]{0.45 \textwidth}
\epsfxsize=\boxsize
\epsfysize=2.2in
\epsfbox{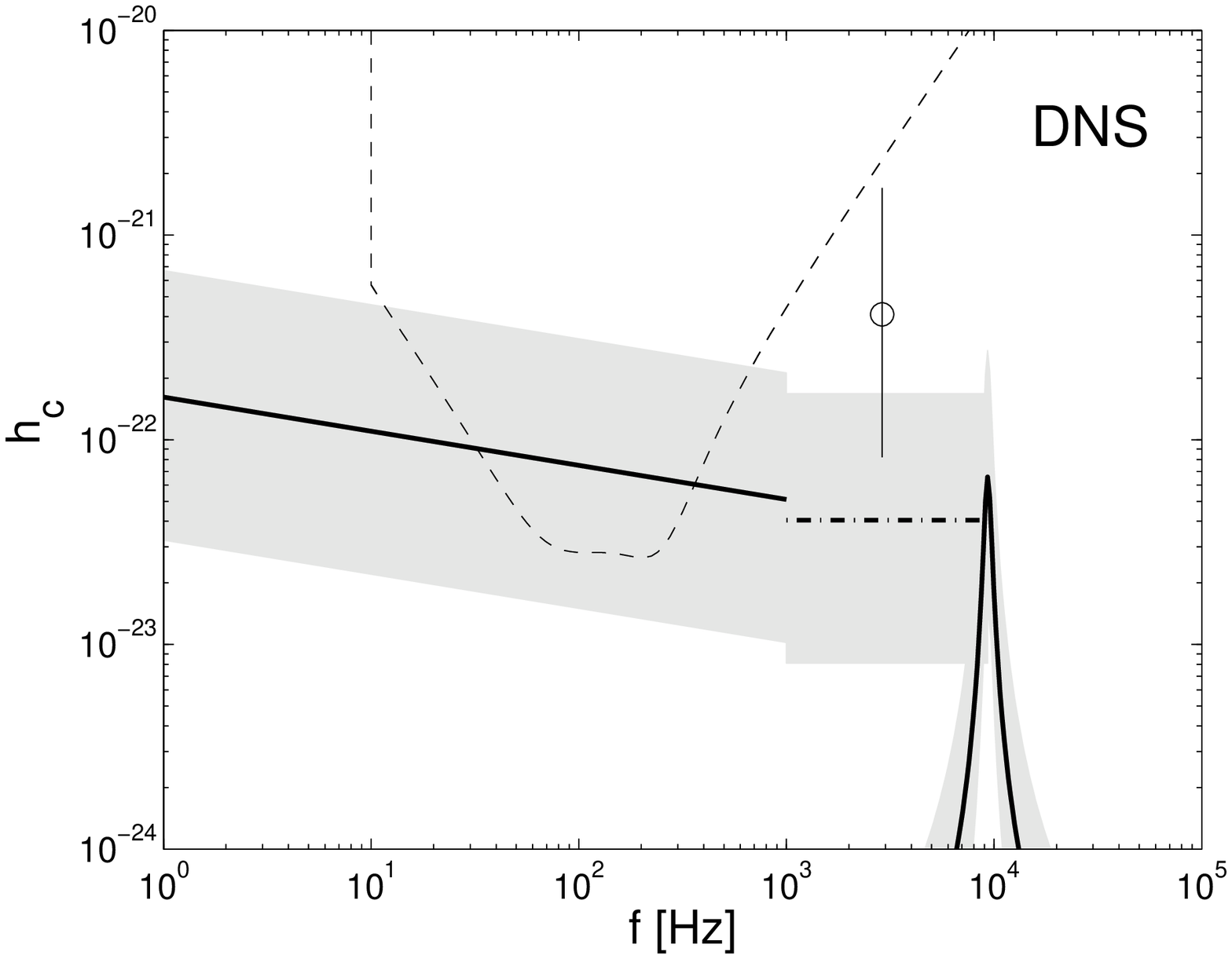}
\end{minipage}
\hspace{3mm}
\begin{minipage}[t]{0.45\textwidth}
\epsfxsize=\boxsize
\epsfysize=2.2in
\epsfbox{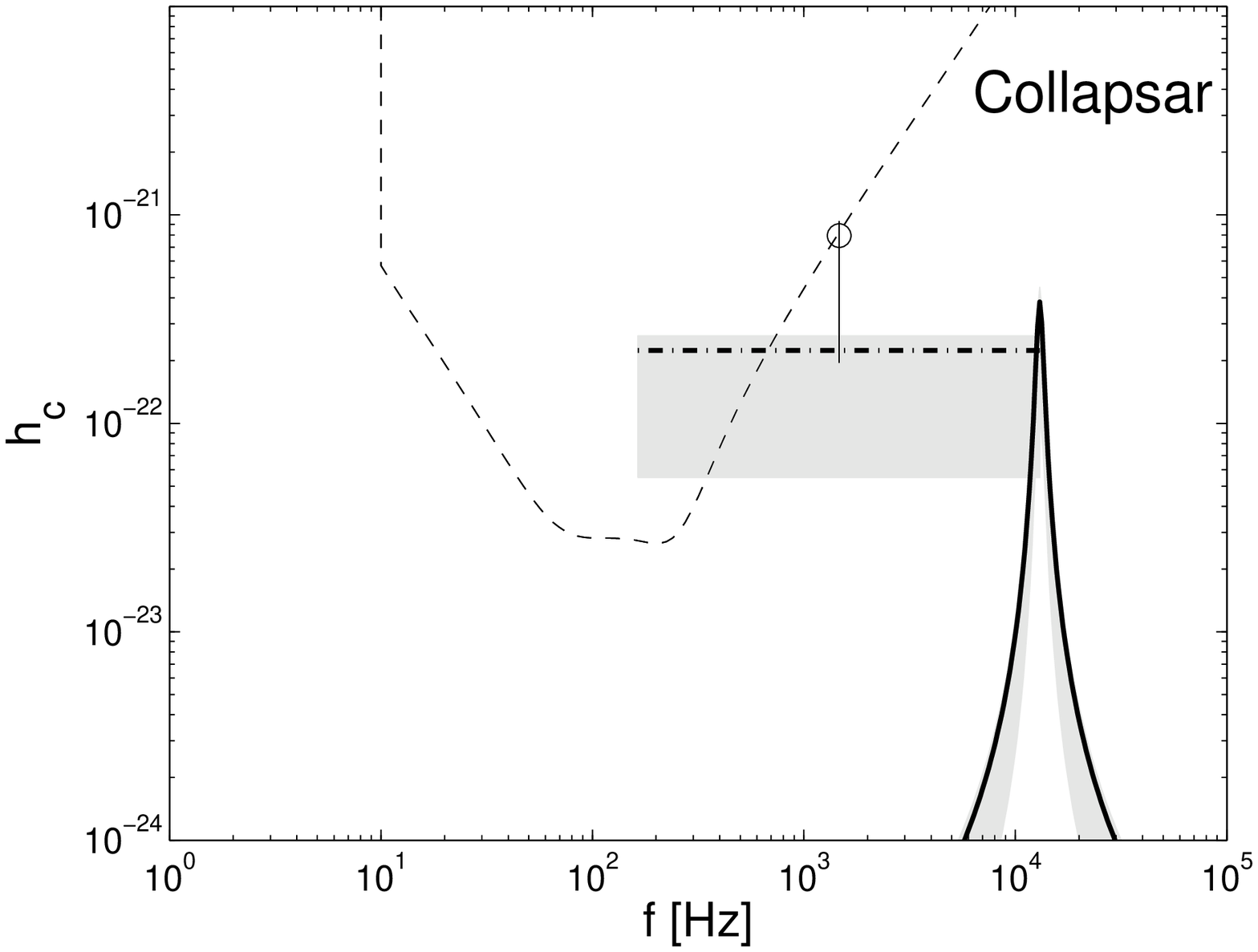}
\end{minipage}
\caption{{\it Left panel:} Characteristic GW strains for merging double neutron 
stars : in-spiral (solid line), merger (dashed dotted line), bar (circle), 
ring-down(solid spike).
Also shown is the advanced LIGO nose curve $\sqrt{f S_h(f)}$ (dashed curve).
The shaded region and the vertical line reflect the uncertainty of the
formation rate $R$ in Table 1.
{\it Right panel}: Characteristic GW strains for collapsars: blob merger 
(dashed dotted line), bar (circle) and ring-down(solid spike).}
 \label{fig:dns}
\end{center}
\end{figure}
%%%%%%%%%%%%%%%%%%%%%%%%%%%%%%%%%%%%%%%%%%%%%%%%%%%%%%%%%%%%%%%%%%%%%%%%%%%

For the massive rotating stellar collapse (collapsar) scenario of GRB, the
non-axisymmetrical perturbations may be stronger \cite{davies02,fryer02,vanput02},
and the estimated formation rates are much higher than for other progenitors
\cite{fryer02,belczynski02}, with typical distances correspondingly much
nearer to Earth. This type of progenitor is of special interest, since it
has so far received the most observational support from GRB afterglow
observations. For collapsars, in the absence of detailed numerical 3D
calculations specifically aimed at GRB progenitors, we have estimated \cite{km03}
the strongest signals that might be expected in the case of bar instabilities
occurring in the accretion disk around the resulting black hole, and in the
maximal version of the recently proposed fragmentation scenario of the
infalling core. Although the waveforms of the gravitational waves produced in
the break-up, merger and/or bar instability phase of collapsars are not known,
a cross-correlation technique can be used making use of two co-aligned detectors.
Under these assumptions, collapsar GRB models would be expected to be marginally
detectable as gravitational wave sources by the advanced LIGO within one year
of observations.

In the case of binaries the matched filtering technique can be used, while
for sources such as collapsars, where the wave forms are uncertain, the
simultaneous detection by two elements of a gravitational wave interferometer,
coupled with electromagnetic simultaneous detection, provides a possible
detection technique.  We have made \cite{km03} more specific detection estimates
for both the compact binary scenarios and the collapsar scenarios,

Both the compact merger and the collapsar models have in common
a high angular rotation rate, and observations provide evidence
for jet collimation of the photon emission, with properties
depending on the polar angle, which may also be of relevance for
X-ray flashes. We have considered \cite{km03b} the gravitational wave
emission and its polarization as a function of angle which is expected
from such sources.  The GRB progenitors emit $l=m=2$ gravitational
waves, which are circularly polarized on the polar axis, while the $+$
polarization dominates on the equatorial plane. Recent GRB studies suggest
that the wide variation in the apparent luminosity of GRBs are caused by
differences in the viewing angle, or possibly also in the jet opening angle.
Since GRB jets are launched along the polar axis of GRB progenitors,
correlations among the apparent luminosity of GRBs ($L_\gamma(\theta)\propto
\theta^{-2}$ and the amplitude as well as the degree of linear polarization $P$ 
degree of the gravitational waves are expected, $P\propto \theta^{4}\propto 
L_\gamma^{-2}$.
At a viewing angle larger than the jet opening angle $\theta_j$
the GRB $\gamma$-ray emission may not be detected. However, in such cases
an ``orphan'' long-wavelength afterglow could be observed, which would be
preceded by a pulse of gravitational waves with a significant linearly
polarized component. 
%An expanding jet with an opening angle $\theta_j$
%behaves, as long as its Lorentz factor  $\gamma > \theta_j^{-1}$, as if it
%were part of a spherical shell, but relativistic beaming effect allows only
%observers at viewing angles $< \theta_j$ to observe the emission from the jet.
As the jet slows down and reaches $\gamma \sim \theta_j^{-1}$, the jet begins
to expand laterally, and its electromagnetic radiation begins to be observable
over increasingly wider viewing angles. Since the opening angle increases as
$\sim\gamma^{-1} \propto t^{1/2}$,  at a viewing angle $\theta > \theta_j$,
the orphan afterglow begins to be observed (or peaks) at a time $t_{p}
\propto \theta^2$ after the detection of the gravitational wave burst.
The polarization degree and the peak time should be correlated as
$P\propto t_p^2$.

A new type of fast transient source, called ``X-ray flashes'', have
recently been observed with the Beppo-SAX satellite \cite{kippen02}.
Apart from their large fraction of X-rays ($\sim 2-10$ keV), the overall
properties of these events are similar to those of GRBs. The nature of
these sources is  not yet fully understood. Although they could be a totally
different astrophysical phenomenon, it has been suggested that these events
may be GRBs with large viewing angles \cite{yamazaki,wzh02}. If this
is the case, linearly polarized
gravitational waves should be observed prior to the X-ray flashes. The
degree of polarization should be positively correlated with longer delays
and with the softness of the X-ray flashes, which increase with angle.
Since the degrees of linear and circular polarization depend on the
viewing angle, a determination of the polarization degree would be a
measure of the viewing angle. Such measurements, which are likely to
require the advent of a future generation of detectors, could provide
a new a tool for estimating the absolute luminosity of GRBs, including
its photon component. By comparing the estimated absolute photon
luminosity with the apparent luminosity, the distance to the source may
be estimated independently of any redshift measurement. No optical
afterglows have been found for about half of all the GRBs detected by
Beppo-SAX (the so called ``dark GRBs''), and the present method would
have the potential to help determine or constrain the distances to such
dark GRBs.

\bigskip
PM is grateful to the Niels Bohr Institute and Nordita as well as the symposium
organizers for their kind hospitality, and to NSF AST-0098416, NASA NAG5-9192
and 9153 for support.

%%%%%%%%%%%%%%%%%%%%%%%%%%%

\def\Discussion{
\setlength{\parskip}{0.3cm}\setlength{\parindent}{0.0cm}
     \bigskip\bigskip      {\Large {\bf Discussion}} \bigskip}
\def\speaker#1{{\bf #1:}\ }
\def\endDiscussion{}

\Discussion

\speaker{A. Beloborodov (Columbia University)} 
An additional mechanism for the high-energy break in the gamma-gamma
absorption is the radiation scattered off the ambient medium. For WR-wind medium
one expects a break at 10-50 MeV.

\speaker{\Mesz} Yes, that is a potentially important effect, which you have discussed.

\speaker{M.J. Rees (Cambridge University)}
The Bahcall/Waxman limit  could perhaps be violated if ions were accelerated
in a confined and opaque region, and didn't  escape to contribute to the cosmic 
ray flux. I mention this just to reduce pessimism!

\speaker{\Mesz} I agree, sources which are optically thick to CRs could
violate the WB limit.

\speaker{T. Piran (Hebrew University)} 
Gravitational waves can also arise from the acceleration process of the
matter in GRBs.

\speaker{\Mesz} This is an interesting effect, which as you discussed, would have 
a significantly lower flux than the main burst from mergers or core collapse.

\endDiscussion

\end{document}

%% file: econfmacros.tex
\def\beq{\begin{equation}}
\def\eeq#1{\label{#1}\end{equation}}
\def\eeqn{\end{equation}}

\def\beqa{\begin{eqnarray}}
\def\eeqa#1{\label{#1}\end{eqnarray}}
\def\eeqan{\end{eqnarray}}

\let\bar=\overbar

\def\Dslash{\not{\hbox{\kern-4pt $D$}}}
\def\dslash{\not{\hbox{\kern-2pt $\del$}}}

\def\msb{{\bar{\ssstyle M \kern -1pt S}}}

\def\eps{\epsilon}